# PIP-II SSR2 CAVITIES FABRICATION AND PROCESSING EXPERIENCE*

M. Parise†, D. Passarelli, P. Berrutti  
Fermi National Accelerator Laboratory (Fermilab), 60510 Batavia, IL, USA  
D. Longuevergne, P. Duchesne  
Irène Joliot-Curie Laboratoire (IJCLab), CNRS-IN2P3, 91406, Orsay, France

*Abstract*

The Proton Improvement Plan-II (PIP-II [1]) linac will include 35 Single Spoke Resonators type 2 (SSR2). A pre-production SSR2 cryomodule will contain 5 jacketed cavities. Several units are already manufactured and prepared for cold testing. In this work, data collected from the fabrication, processing and preparation of the cavities will be presented and the improvements implemented after the completion of the first unit will be highlighted.

## INTRODUCTION

The Radio-Frequency (RF) design and the mechanical design of the pre-production SSR2 cavities was presented [2], [3]. The acquisition of the jacketed cavities, which were set to be produced by the industrial sector, was launched in early 2021, resulting in the first bare unit being finished by the year's end. Following a bulk Buffered Chemical Polishing (BCP), a thorough High Pressure Rinse (HPR) of the RF space, and a High Temperature Heat Treatment (HTHT), the jacketing process and testing at room temperature of the first cavity were finalized in July 2022. By September 2022 the first unit also received light BCP, HPR, the RF volume was sealed with flanges (2 of which included the antennas for the unity coupler and the field probe) in a cleanroom environment and 120C bake was the last step before shipping the unit. Other 2 units were manufactured by the same supplier and went through the same processing recipe since then. Moreover, the bare cavities' components for the next set of 3 units were formed and welded and they are waiting for the final Electron Beam (EB) welds to complete other 3 bare cavities. Some of the main parameters used for the design of the SSR2 cavities and that will be recalled in this work are reported in Table 1.

Table 1: Parameters of the SSR2 Cavity.

| Parameter | Value |
| --- | --- |
| Nominal Frequency, MHz | 325 |
| df/dp, $\frac{Hz}{mbar}$ | <25 |
| Target Frequency Allowable Error, kHz | +/-50 |
| Maximum Allowable Working Pressure (MAWP) RT / 2 K, bar | 2.05 / 4.1 |

* WORK SUPPORTED BY FERMI RESEARCH ALLIANCE, LLC UNDER CONTRACT NO. DEAC02- 07CH11359 WITH THE UNITED STATES DEPARTMENT OF ENERGY  
† mparise@fnal.gov

## NIOBIUM FORMING

The crucial components to be fabricated include the endwalls, spokes, and spoke collars. In order to test the capabilities of the forming technique and equipment in producing components within the required mechanical tolerances, copper sheets are used as a substitute for precious niobium. Once the process is verified to yield the expected outcomes, the same procedures are followed to acquire the final components.

Metal spinning is employed to fabricate the endwalls, while deep drawing is utilized to obtain the spokes and spoke collars. The endwalls undergo the most significant reduction in thickness during the spinning process.

The thickness post spinning or forming is measured at (see Fig. (1)):

- 36 different locations: 9 points spaced by 90 degrees for the endwalls;
- 30 different locations for the half spokes;
- 16 different locations: 4 points spaced by 90 degrees for the collars.

and the results are reported in Fig. 2, 3 and 4 as an average percentage thickness reduction at each location.

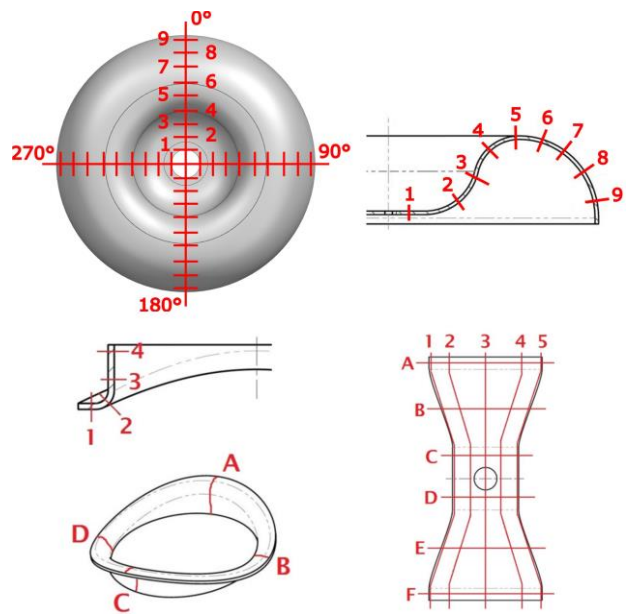

Figure 1: Points' Location where the Thickness is Measured.

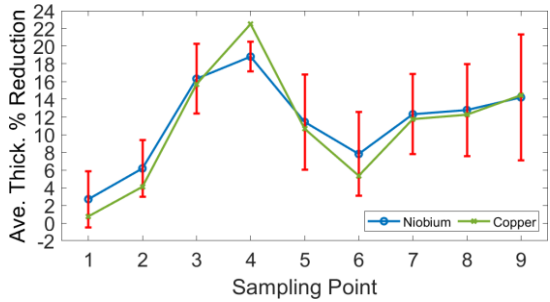

Figure 2: Average Percentage Thickness Reduction on the Endwalls used to Fabricate the first 6 SSR2 Cavities.

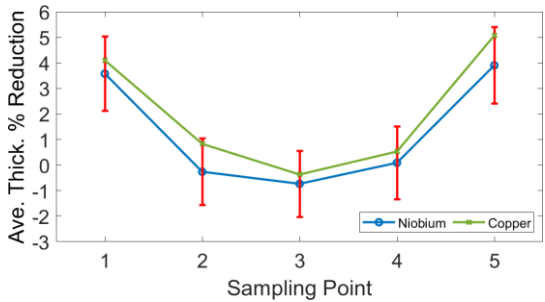

Figure 3: Average Percentage Thickness Reduction on the Half Spokes used to Fabricate the first 6 SSR2 Cavities.

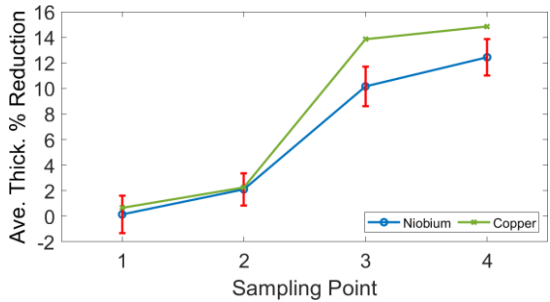

Figure 4: Average Percentage Thickness Reduction on the Collars used to Fabricate the first 6 SSR2 Cavities.

The error bars extremities represent the maximum and minimum values over the 12 endwalls, 12 half spokes and 12 collars respectively.

The plots also show the thickness reduction that was measured for the copper trials (1 copper trial for each shape). Overall, a good agreement between copper and niobium spinning/forming can be observed, confiming the usefulness of the trials as an effective tool to establish the goodness of the forming step.

The maximum average percentage thickness reduction is 20 % in correspondence of the inflection of the endwall.

## FREQUENCY HISTORY THROUGHOUT MANUFACTURING AND PROCESSING

Before the start of the manufacturing and processing activities a target frequency recipe is developed starting from the experience and lessons learned from past projects [5], [6]. The first 3 units were not manufactured in parallel, therefore the frequency recipe was updated based on the frequency shifts measured along the cavities' life-cycle. Fig. 5 shows the frequency of the 3 pre-production SSR2 (ppSSR2) cavities starting from the bare cavity (the results of the frequency trimming were presented [4]) and ending after the last processing step: the low temperature baking.

The greatest uncertainties resided in the frequency shifts due to the High Temperature Heat Treatment (HTHT), which is difficult to forecast due to the stress relief phenomena of the formed niobium, and of the helium vessel integration (Weld Step 1 through 8 in Fig. 5). Although the initial frequency are not aligned due to a manufacturing error during the frequency trimming, the individual frequency shift due to the processing steps such as Buffered Chemical Processing (BCP), High Temperature Heat Treatment (HTHT), Helium vessel weld steps and leak checks are comparable among the 3 cavities.

As pointed out in [4] there is 1 weld during the integration of the helium vessel that cause the greatest of the frequency shifts among the welding steps and it is represented at "Weld Step 2". A new welding sequence was implemented after the completion of the first unit but the frequency shift due to this step did not improve. The average frequency shift resulting from the welds between the titanium jacket and the bare cavity is 130 kHz. The first unit exhibits the highest total shift at 147 kHz, while the remaining two units show individual shifts of 120 kHz each.

Each cavity is pressure tested until the 117 % of the MAWP according to [7] at room temperature (2.4 bar-g). The pressure is gradually incremented and at each step the pressure is held for 1 minute and 10 minutes at the last step while checking for leaks or pressure drops at the minimum scale of the pressure gauge. All cavities successfully passed the pressure test also showing minimal overall frequency shift.

The pressure sensitivity (see Table 1) can be extrapolated from the pressure test frequency measurements and it is confirmed during the cold tests. All 3 units have a sensitivity well within the 25 Hz/mbar with -4.6, -3.1 and -4.1 respectively.

Table 2, shows the frequency shifts during the cavities cold tests. The values show little variations between the 3 units at each step. The last line (Tuner Engaged) represents the frequency adjustment required to bring the cavity the desired frequency of 325.000 MHz.

Table 2: Frequency shifts (kHz) during cold tests

| **Step** | **Cav. 1** | **Cav. 2** | **Cav. 3** |
| --- | --- | --- | --- |
| Insulating Vacuum | 202 | 203 | 242 |
| 4K | 476 | 465 | 460 |
| 2K (He space Vacuum) | -3 | 3 | 4 |
| Tuner Engaged | -21 | -88 | -9 |

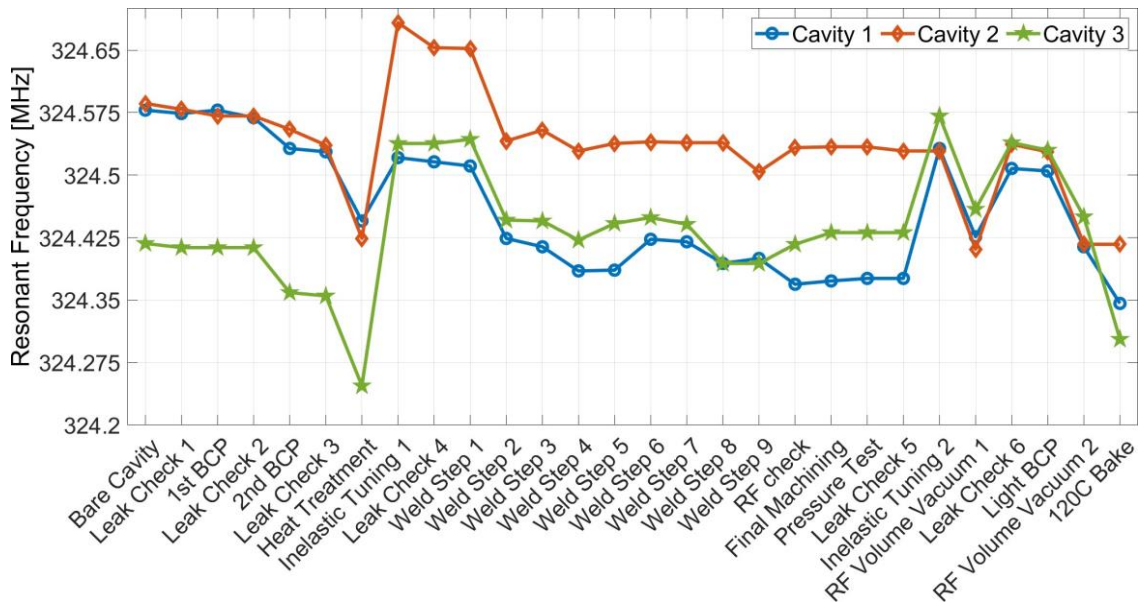

Figure 5: Frequency history

Fig. 6 shows the first unit as received after the tuner installation and prior to the cold test in the test cryostat.

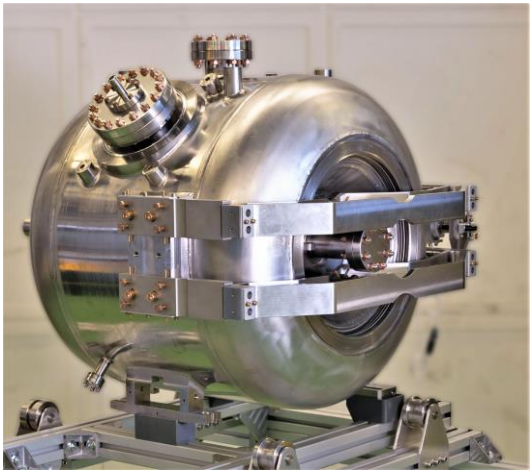

Figure 6: First jacketed cavity to be completed with the tuner installed.

## BUFFERED CHEMICAL POLISHING (BCP)

The rotational BCP treatment, as opposed to a static BCP, is a necessary step during SRF cavity production, which consists of removing the superficial layer of the inner surface of the niobium through a chemical polishing. The SSR2 cavity requires 2 different rotational BCP treatments:

- Bulk BCP
- Light BCP

A total of 2 cavities were processed in industry. The bulk BCP treatment is performed on the bare cavity and the total amount of niobium removed from the cavity inner surface should be between 145 $\mu$ m and 155 m. The light BCP shall be done on the jacketed cavity and the material removed shall be between 25 $\mu$ m and 30 $\mu$ m. The composition of the mixture is 1:1:2 in volume of hydrofluoric, nitric and ortho-phosphoric acids. The reaction rate shall be controlled by keeping the acid temperature below 15°C for all the duration of the treatment to minimize the risk of contaminating the bulk niobium with hydrogen. Both BCP treatments (bulk and light) are performed rotating the cavity around the side port axis as shown Fig. 7.

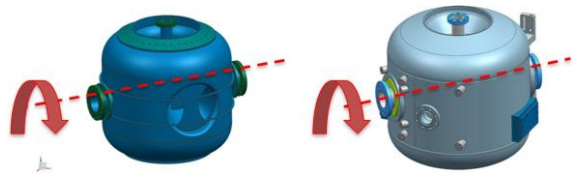

Figure 7: Bare cavity and jacketed cavity rotation axis during both bulk and light BCP treatments.

Fig. 8 depicts the 28 precise locations on the exposed cavity surface where an ultrasonic probe is employed to measure the thickness both before and after the treatment. The average material removed on the 2 cavities processed in industry are 146 $\mu$m and 145 $\mu$m respectively, both with a standard deviation of 12 $\mu$m. The third cavity that was processed at IJCLab went through a difference sequence of rotational bulk BCPs with an average total thickness reduction of 166.8 $\mu$m and an overall standard deviation of 14.3 $\mu$m [8].

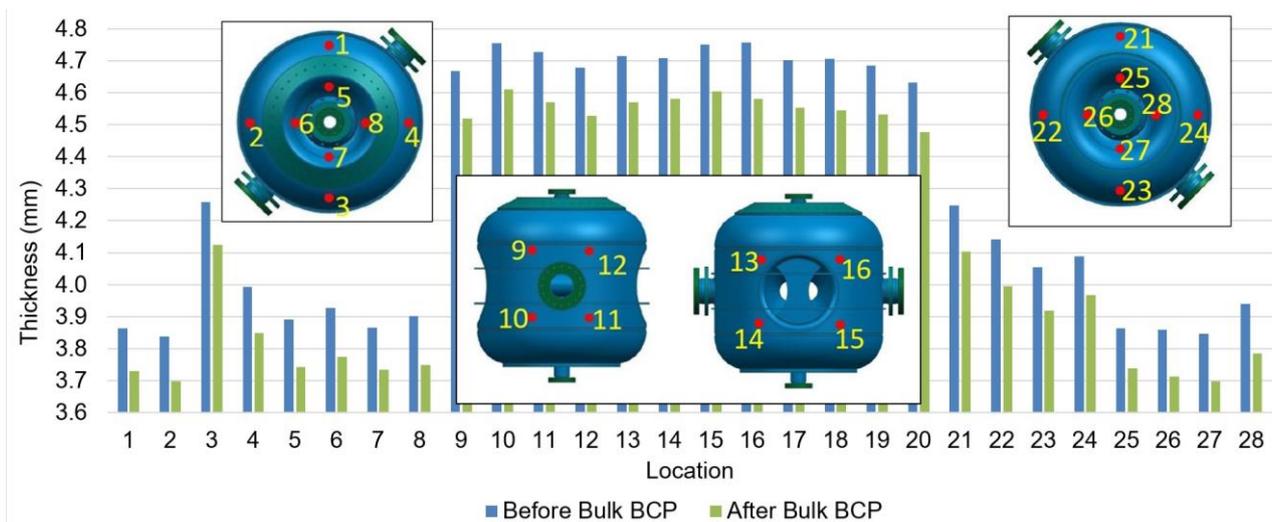

Figure 8: Thicknesses before and after bulk BCP.

During the BCPs the cavity is filled with acid at a temperature of 2°C, until it reaches 60% of its internal volume. Throughout the treatment process, a nitrogen flow is introduced into the cavity to minimize the generation of gases resulting from the reaction. Simultaneously, a scrubber is employed to continuously remove these gases. The external surface temperatures of the cavity are monitored using a set of eight temperature sensors. The temperature of the acid is monitored at both the entrance and the return points. Fig. 9 shows the trend of the acid inlet and outlet temperature during a bulk BCP as well as the cavity temperatures.

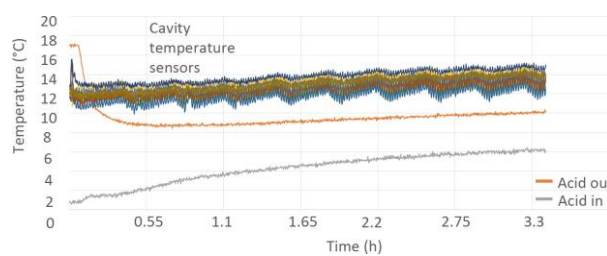

Figure 9: Acid and cavity temperature during a bulk BCP.

## HIGH PRESSURE RINSE

During the RF cold test of the first unit, the cavity performance was degraded by Field Emission (FE). This motivated an in-depth investigation of the cavity manufacturing and processing steps. While the causes of the FE are not fully understood, this investigation led to the development of an analytical tool capable of showing the areas rinsed by a particular HPR system given the geometry of the cavity, the nozzle head (orientation and number of the nozzles), water spray angle, rotational and linear wand speeds and penetration of the wand into each cavity port.

This tools was applied to 3 different HPR setups (at the vendor, at IJCLab and at Fermilab) for 2 nozzle head ge-ometries. The vendor's and IJCLab's HPR systems have the wand oriented vertically while Fermilab system is horizontal.

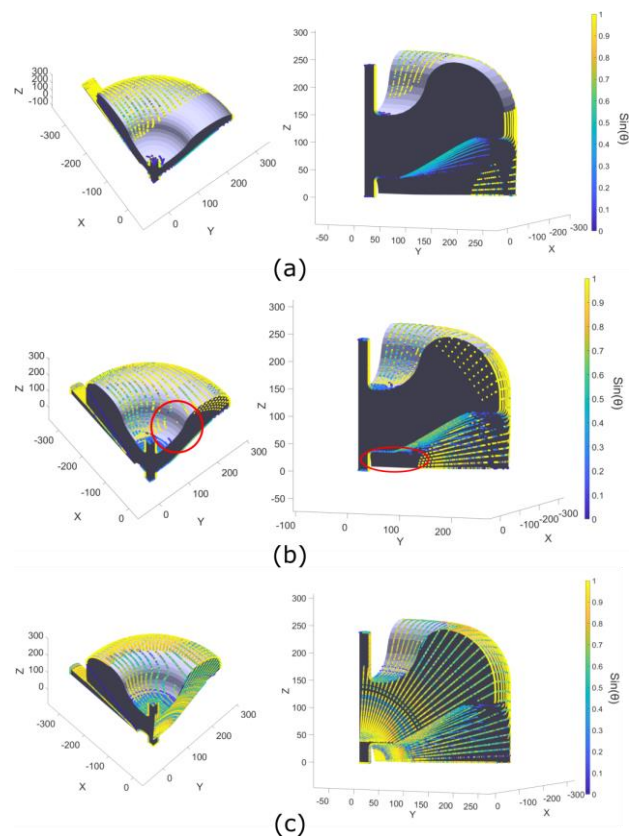

Figure 10: HPR Simulations decipting the areas rinsed by different HPR systems. (a) Rinsed areas for the first unit, clearly shows blindspots. (b) Rinsed areas with Fermilab's horizontal HPR system, circled in red areas with high electric field not rinsed. (c) Simulation of rinsed areas expected from improved HPR nozzle head that will be used at Fermilab.

Fig. 10 shows the results of this analysis. It is clear from subset a that some areas of the cavity were not properly rinsed, including high Electric Field regions which may give rise to FE. Based on the results 2 new nozzle head geometries were developed for the IJCLab and Fermilab setups in order to improve the wetted surface and the efficacy of the rinsing (which is proportional to the sin of the angle of attack of the water). These 2 nozzles are currently being procured and will used in the next HPR cycles.

Additionally, both laboratories are implementing HPR setups with the ability of tilting the wand with respect to the cavity port (see. Fig. 11).This will enable to rinse hard-to-reach areas of the RF volume.

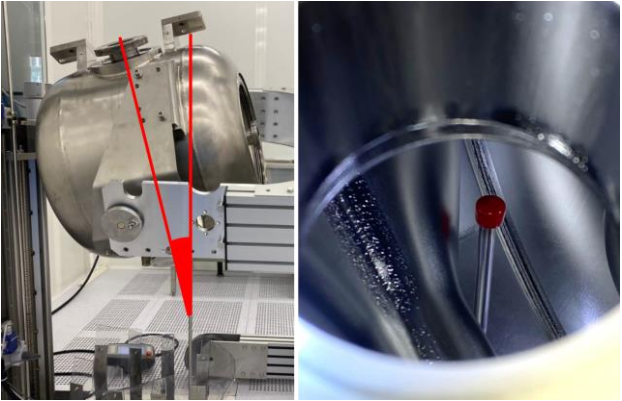

Figure 11: By inserting the wand into the cavity side port with an angle, hard-to-reach areas are better rinsed.

Fermilab is also evaluating the possibility of performing 2 subsquent HPRs with the first in a horizontal wand configuration and the second with a vertical wand. Due to the shape of the cavity endwall, vertical HPR may be ineffective if the wand does not protrude beyond the spoke pipe, which was the case during the first iterations.

Cold tests repeated after the implementation of the new HPR setups will be able to quantify the performance improvements introduced with the changes described above.

## CONCLUSION

A total of 3 ppSSR2 jacketed cavity are manufactured and processed and others are soon to be completed. The cavities are within the frequency target range and passed the pressure tests. This demonstrates a solid mechanical design and repeatable manufacturing process. The BCP data show uniform material removal across the cavity RF surface. HPR nozzle configuration and parameters were studied showing that the current setups both at the vendor and the laboratories can be greatly improved. New HPR's nozzle head designs were finalized and are soon to be implemented.